\documentclass[prl,twocolumn,showpacs,floatfix,superscriptaddress,citeautoscript]{revtex4}

\usepackage{graphicx}

\begin{document}

\title{Asymmetric weak-pinning superconducting channels: vortex ratchets}

\author{K. Yu}
\affiliation{Department of Physics, Syracuse University, Syracuse, New York 13244-1130, USA}
\author{T. W. Heitmann}
\affiliation{Department of Physics, Syracuse University, Syracuse, New York 13244-1130, USA}
\author{C. Song}
\affiliation{Department of Physics, Syracuse University, Syracuse, New York 13244-1130, USA}
\author{M. P. DeFeo}
\affiliation{Department of Physics, Syracuse University, Syracuse, New York 13244-1130, USA}
\author{B. L. T. Plourde}
\email[]{bplourde@phy.syr.edu}
\affiliation{Department of Physics, Syracuse University, Syracuse, New York 13244-1130, USA}
\author{M. B. S. Hesselberth}
\affiliation{Kamerlingh Onnes Laboratorium, Leiden University, P.O. Box 9504, 2300 RA Leiden, The Netherlands}
\author{P. H. Kes}
\affiliation{Kamerlingh Onnes Laboratorium, Leiden University, P.O. Box 9504, 2300 RA Leiden, The Netherlands}

\date{\today}


\begin{abstract}
The controlled motion of objects through narrow channels is important in many fields.
We have fabricated asymmetric weak-pinning channels in a superconducting thin-film strip for controlling the dynamics of vortices. The lack of pinning allows the vortices to move through the channels with the dominant interaction determined by the shape of the 
channel walls. 
We present measurements of vortex dynamics in the channels and compare these with similar measurements on a set of uniform-width channels. While the uniform-width channels exhibit a symmetric response for both directions through the channel, the vortex motion through the asymmetric channels is quite different, with substantial asymmetries in both the static depinning and dynamic flux flow. This vortex ratchet effect has a rich dependence on magnetic field and driving force amplitude.
  \end{abstract}

\maketitle

Recently there has been much interest in developing artificial ratchets for generating directed motion using tailored asymmetries \cite{Hanggi05}. Such ratchets could be used as pathways for producing net transport of matter at the nanoscale. In addition, artificial ratchets can serve as model systems for understanding similar ratchet phenomena in biological systems while allowing for experimental control over many of the ratchet parameters \cite{Astumian97}. A variety of ratchets have been considered, but one particular type that has been implemented in several different systems is the rocking ratchet, where a spatial asymmetry is engineered into the potential energy landscape governing particle motion and an external control variable can be adjusted to tilt this potential. The application of an oscillatory drive of the control variable with zero mean can result in the net motion of particles through the potential because of the different rates for overcoming the barriers in the two directions through the ratchet.
  
Implementations of ratchets in solid-state devices include asymmetric structures of electrostatic gates above a two-dimensional electron gas \cite{linke99}, and arrays of Josephson junctions with asymmetric critical currents \cite{shalom05}. Structures have also been developed for producing a ratchet effect with vortices in superconducting thin films involving either asymmetric arrangements of pinning centers \cite{vondel05,togawa05} or asymmetric magnetic pinning structures \cite{villegas03}.
In this Communication, we describe a vortex ratchet using two-dimensional guides to generate asymmetric channels for vortex motion. In our structures, the potential asymmetries arise from differences in the interaction strength between vortices and the channel walls, resulting in a substantial ratchet effect for the motion of vortices through the channels.
Our design is related to a previous vortex ratchet proposal \cite{wambaugh99}, although our ratchet is in a somewhat different parameter regime.

Nanoscale channels for guiding vortices through superconducting films with a minimal influence from pinning have been developed for studies of vortex matter in confined geometries, including experiments on melting \cite{besseling03}, commensurability \cite{pruymboom88}, and mode locking \cite{kokubo02}. 
Such channels are fabricated from bilayer films of amorphous-NbGe, an extremely weak-pinning superconductor, and NbN, with relatively strong pinning. A reactive ion etching process removes NbN from regions as narrow as $100$~nm, defined with electron-beam lithography, to produce weak-pinning channels for vortices to move through easily. In contrast, vortices trapped in the NbN banks outside of the channels remain strongly pinned. 

\begin{figure}
\centering
  \includegraphics[width=3.35in]{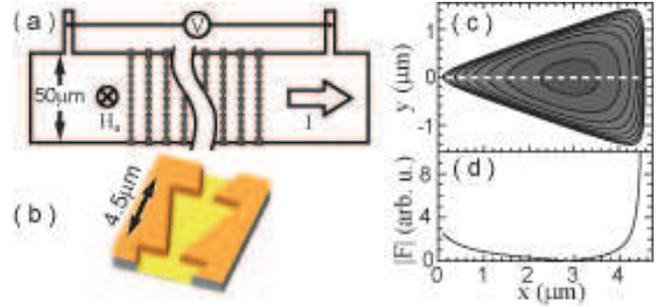}
  \caption{(Color online) (a) Schematic of strip with ratchet channels; channel spacing is $10$~$\mu$m. (b) AFM image of one ratchet cell; channel depth is $88$~nm. (c) Contour plot of model potential for vortex interacting with ratchet cell walls. (d) Magnitude of corresponding force along center of channel.
\label{fig:schematic}}
\end{figure}

We have fabricated weak-pinning channels with $200$~nm-thick films of a-NbGe and $50$~nm-thick films of NbN on a Si substrate, and we have designed many of the channels such that the walls have an asymmetric sawtooth pattern (Fig. \ref{fig:schematic}). Our layout consists of a strip with multiple pairs of probes for sensing the voltage drop $V$ due to vortex motion. A transport current driven through the strip with an external supply generates a transverse Lorentz force on the vortices.
Between each pair of voltage probes is an array of identical channels -- one array consists of $50$ channels each with a constant width of  $2~\mu$m; another array has $30$ ratchet channels with the dimensions described in Figure \ref{fig:schematic}; yet another array contains $30$ identical ratchet channels, all oriented in the opposite direction across the strip. 
We perform our measurements with the strip immersed in a pumped helium bath with a temperature stability of $0.2$~mK/hr. 
Our results presented here were obtained at $T = 2.78$~K, and our measured transition temperature for the a-NbGe is $T_c = 2.88$~K. 
For each measurement sequence, the strip was heated to $\sim 15$~K, above $T_c$ of both the NbGe and NbN films, and was then cooled in zero applied magnetic field; a $\mu$-metal shield reduced the background magnetic field below 13 mG. All field-dependence data were acquired while increasing the magnetic field $H_a$ from zero, where we generate $H_a$ with a superconducting coil.

One can expect such asymmetric channel structures to influence vortex dynamics if the confinement from the channel walls distorts the screening currents that circulate around each vortex differently depending on the motion of the vortex through the channel.
At the temperature of our measurements, we estimate the penetration depth of the NbN to be $\lambda_{NbN} \approx 0.5$~$\mu$m and that of NbGe to be $\lambda_{NbGe} \approx 1.9$~$\mu$m, based on the film parameters and the standard dirty-limit expressions and assuming a two-fluid model for the temperature dependence. Furthermore, the thin-film penetration length, $2 \lambda^2/d$, that sets the characteristic extent for the screening currents around a vortex in a thin film is $\sim 42$~$\mu$m for the NbGe in the channels, clearly much greater than the width of the channels, such that the shape of the channel walls will play an important role in distorting each vortex.
The interaction of a vortex with the channel walls can be understood by considering the model of Mkrtchyan {\it et al.} for the interaction between 
a vortex and the interface between two superconductors with different penetration depths \cite{mkrtchyan72}. 
For our strips, the channel corresponds to the superconductor with the larger penetration depth, while the NbN banks have the shorter penetration depth. 
According to the Mkrtchyan model, a vortex in the channel will experience a repulsive interaction $U_i$ from the $i^{th}$ wall a distance $d_i$ away keeping the vortex in the channel, 
\begin{equation}
 U_i \propto \left( \frac{\lambda_{NbGe}^2-\lambda_{NbN}^2}{\lambda_{NbGe}^2+\lambda_{NbN}^2}\right) \ln \left( \frac{\lambda_{NbGe}}{d_i} \right).
\label{eq:mkrtchyan-pot}
\end{equation}

If we consider a single vortex located in one of the ratchet cells, we can make a crude model of the potential energy landscape by summing the contributions from the interaction of the vortex with each of the three walls of the ratchet cell, $\Sigma U_i$ [Fig. \ref{fig:schematic}(c)]. 
The derivative of this potential along the central symmetry line of the cell exhibits an asymmetric force on the vortices [Fig. \ref{fig:schematic}(d)]. Thus, the two sloped walls result in a gradual increase in the potential energy as the vortex approaches the aperture in the ``easy" direction, while the potential energy grows abruptly as the vortex approaches the wide back wall of the ratchet cell for motion in the ``hard" direction.

\begin{figure}
\centering
  \includegraphics[width=3.35in]{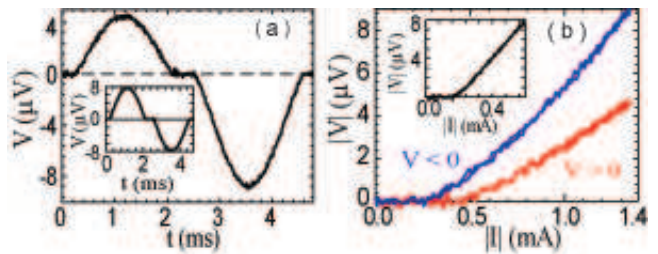}
  \caption{(Color online) (a) $V(t)$ for ratchet channels with sinusoidal current drive $I(t)$; $I_{ac}=1.35$~mA. (Inset) $V(t)$ for $2$~$\mu$m-wide uniform channels; $I_{ac}=0.59$~mA. (b) IVC for ratchet plotted with positive and negative branches in first quadrant for comparison. (Inset)  corresponding IVC for $2$~$\mu$m-wide uniform channels, also with both branches in first quadrant. $H_a = 4.20$~Oe in all plots.
\label{fig:ratchet-IV}}
\end{figure}

The vortex dynamics in the channels can be characterized by measuring $V$, which is proportional to the vortex velocity and density. 
We measure $V$ with a room-temperature amplifier and we drive the vortices by applying $200$ cycles of a bias current sine wave $I(t)$ at $210$~Hz with amplitude $I_{ac}$. We average the resulting voltage response to obtain a $V(t)$ curve for one period. For the uniform channels, this is always symmetric about $V = 0$, while for the ratchets, one side of the curve typically has a larger response than the other [Fig. \ref{fig:ratchet-IV}(a)].  We combine this resulting $V(t)$ curve with $I(t)$ to obtain a current-voltage characteristic, IVC. By plotting the negative and positive branches of the IVC both in the first quadrant, the substantial asymmetry of the response for the ratchet channels is apparent, while the corresponding IVC for the uniform-width channels is symmetric [Fig. \ref{fig:ratchet-IV}(b)]. Furthermore, from the IVC for the ratchet channels, there are clear asymmetries both in the critical currents at which the vortices begin to depin from the static state and in the flux flow resistances, inversely related to the vortex dynamic friction in the channels.

\begin{figure}
\centering
  \includegraphics[width=3.35in]{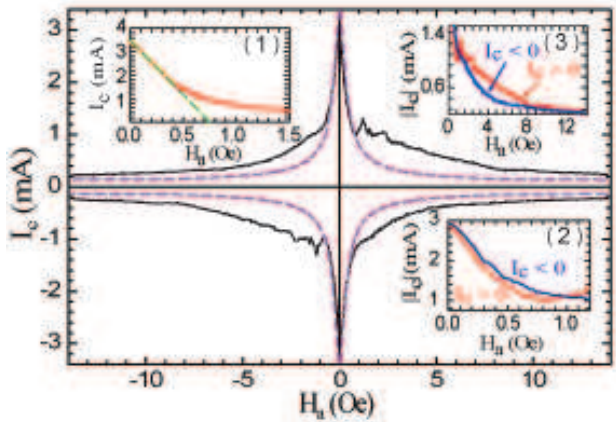}
  \caption{(Color online) Critical current variation with $H_a$ -- ratchet channels (black) and uniform-width channels (magenta, dashed). (Inset 1) $I_c(H_a)$ for uniform-width channels at low magnetic fields along with linear fit. Ratchet $I_c(H_a)$ for both senses of $I$ at (Inset 2) small $H_a$,  (Inset 3) large $H_a$.
\label{fig:critical-current}}
\end{figure}

We characterize the transition from the static state to a dynamical flux flow regime by measuring the critical current in the conventional way, that is, by measuring the IVC as described earlier, then using a $1~\mu$V criterion to define the critical current $I_c$. 
Measurements of the field dependence $I_c(H_a)$ on the $2$~$\mu$m-wide uniform channels display a similar response to that characteristic of an edge barrier for a thin, weak-pinning superconducting strip in a perpendicular magnetic field, where the entry of vortices at the strip edge is determined by the distortion of the current density across the width of the strip  \cite{plourde01, vodolazov00}.
$I_c$ is a maximum at $H_a = 0$, where $I_c$ is due to the entry of vortices and antivortices at opposite edges of the strip due to the self-field of $I$.
As $H_a$ is increased, $I_c(H_a)$ initially decreases linearly, as the self-field and $H_a$ add with the same sense at one edge and are able to exceed the vortex entry condition for progressively smaller $I$. In this regime, there are no vortices present in the strip for $I < I_c$, while larger currents result in a dynamical flux flow state with vortices entering the strip at one edge and moving across to the other edge.
For larger $H_a$, the external field can be sufficient to push vortices into the strip, even for $I = 0$, and these vortices arrange in a static dome-shaped structure in the middle of the strip \cite{vodolazov00, benkraouda98}. When $I \neq 0$, the dome shifts towards one edge and $I_c$ is reached when the self-field plus $H_a$ at the opposite edge overcome the entry barrier to allow new vortices to enter. In this regime, $I_c$ decreases like $H_a^{-1}$ \cite{plourde01}.
The measurements of $I_c(H_a)$ for the $2$~$\mu$m-wide uniform channels follow essentially this behavior and $I_c$ is symmetric with the direction of $I$ and the sense of $H_a$ (Fig. \ref{fig:critical-current}), indicating that the channels are symmetric and the strip edges at the ends of the channels do not have any significant roughness asymmetries.
This is consistent with the entry of vortices only into the channels at the edge of the strip, and not into the strong-pinning NbN banks, as one would expect at the relatively small magnetic fields of our experiment, based on the lower edge barriers at the channel edges compared to the thicker NbN banks.

For the ratchet channels, $I_c$ is also a maximum at $H_a = 0$, with an initial linear decrease as $H_a$ is increased, however, in contrast to the uniform-width channels, $I_c$ is weakly asymmetric with a $\sim 15 \%$ difference for the two polarities of $I$. In this low-field regime, where $I_c$ corresponds to the entry of vortices into the vortex-free state of the channels, the smaller $I_c$ has the sense of the bias current pushing the vortices in the hard direction of the ratchet. When $H_a$ is reversed, the sense with the smaller $I_c$ inverts as well, again corresponding to vortex motion in the hard direction.
For larger magnetic fields, $I_c(H_a)$ deviates from a linear decrease, as in the uniform-width channel measurements when a static vortex dome can be formed in the channels before $I$ reaches $I_c$.
However, in this regime the ratchet channel $I_c$ develops a substantial asymmetry with respect to the sense of $I$. Now $I_c$ for the sense of $I$ pushing vortices in the hard direction 
becomes considerably larger than that for the easy direction and exhibits a sequence of peaks.
Thus, at the start of this regime, the critical currents for the two directions of vortex motion actually cross [Fig. \ref{fig:critical-current}(inset 2)].
In this regime, vortices sit statically in the ratchet channels for $I < I_c$, and thus can explore the asymmetry due to the shape of the ratchet channel walls as $I$ is increased [Fig. \ref{fig:schematic}], such that vortices depin and flow at a smaller $I$ when the Lorentz force is oriented in the easy direction. This is consistent with the antisymmetry of $I_c(H_a)$, that is, for the opposite sense of $H_a$, the smaller $I_c$ occurs for the opposite sense of $I$, and thus the same spatial direction through the ratchet.
We observe an identical response, but with the opposite sign for ratchet channels with the same geometry but the opposite direction of ratchet cells at a different location on the same strip (not shown).

The abrupt crossing of $I_c$ for the two senses of current corresponds to the first entry of a vortex into each ratchet channel for $I < I_c$ and the peak structure in $I_c$ for the hard direction of the ratchet is likely due to the entry of subsequent vortices into each ratchet channel. 
For small $H_a$, below this crossover of the two senses of $I_c$, where no vortices are present in the channels for $I < I_c$, there are screening currents flowing along the channel walls due to the discontinuity in thickness and penetration depth at each wall. These currents will be concentrated at the outer points of each ratchet cell and can effectively invert the sense of the ratchet potential defined by the shape of the channel walls, thus reversing the ratchet effect for the vortices that enter the channels when $I > I_c$. For larger $H_a$, the interaction of the circulating currents for each vortex in the channel with the walls dominates and the ratchet effect exhibits the sign expected from the spatial asymmetry of the channel pattern. 

\begin{figure}
\centering
  \includegraphics[width=3.35in]{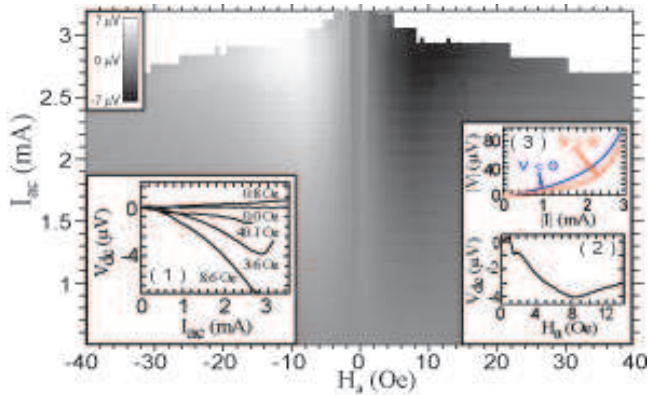}
  \caption{Density plot of $V_{dc}$ {\it vs.} $I_{ac}$ and $H_a$. (Inset 1) Line cuts of $V_{dc}$($I_{ac}$) for indicated values of $H_a$. (Inset 2) $V_{dc}(H_a)$ line cut at $I_{ac}=2.1$~mA. (Inset 3) IVC for $H_a = 8.6$~Oe at measured for large $I_{ac}$.
\label{fig:density-plot}}
\end{figure}

In addition to the asymmetric response of $I_c(H_a)$ for the ratchet channels, we also observe substantial asymmetries in the dynamical flux-flow state. 
A general method to characterize asymmetries in both static and dynamic properties involves averaging $V(t)$ over a complete cycle, such as the trace in Fig. \ref{fig:ratchet-IV}(a), to obtain $V_{dc}$. For a value of $H_a$ corresponding to the IVC of Figure \ref{fig:ratchet-IV} for the ratchet channels, $V_{dc}$ will clearly be non-zero, while for uniform-width channels, we always observe $V_{dc} = 0$ for all $H_a$.
We map the variation of $V_{dc}$ with $H_a$ and $I_{ac}$ for the ratchet channels (Fig. \ref{fig:density-plot}) by zero-field cooling, then measuring $V_{dc}(I_{ac})$ while incrementing $H_a$ towards positive values. We zero-field cool again to measure the $H_a < 0$ response by incrementing $H_a$ from zero towards negative values.
For each $H_a$, we perform our standard measurement of $V_{dc}$ using a burst of sinusoids with amplitude $I_{ac}$ while stepping to progressively larger values of $I_{ac}$.  We continue to increase $I_{ac}$, thus increasing the vortex velocity through the channels, until an instability occurs and the channels switch out to the normal conducting state. The switching point is independent of the frequency of our $I(t)$ sinusoid, at least up to $30$~kHz, and the sample is immersed in liquid helium, thus making simple Joule heating unlikely as the cause. Instead, the curvature in the IVC at large $I_{ac}$ [Fig. \ref{fig:density-plot}(inset 3)] suggests that the switching is related to the Larkin-Ovchinnikov vortex core instability mechanism \cite{larkin75}, perhaps with a related self-heating effect as evidenced by the $H_a$-dependence of the maximum $I_{ac}$ visible in Figure \ref{fig:density-plot} \cite{bezuglyj92}.

For any $H_a$, $V_{dc}(I_{ac})$ is generally zero for small $I_{ac}$, when $I < I_c(H_a)$ for both polarities.
For larger $I_{ac}$, $|V_{dc}|$ tends to grow, and for certain $H_a$, $|V_{dc}|$ eventually begins to decrease before the channels switch out to the normal state [Fig. \ref{fig:density-plot}(inset 1)]. 
In general, $V_{dc}$ is antisymmetric with $H_a$, thus indicating that the direction for net vortex motion corresponds to the same spatial direction through the ratchet channels.
There are also substantial peaks in $|V_{dc}|$ visible on either side of $H_a = 0$, thus there is an $H_a$ that optimizes the ratchet effect (Fig. \ref{fig:density-plot}).
For $H_a=0$, $V_{dc} \approx 0$ for all $I_{ac}$, as there are no screening currents flowing along the channel walls in response to $H_a$.
For $H_a \neq 0$ but small, the sign of $V_{dc}$ corresponds to the net motion of vortices in the hard direction, consistent with the reversal of the critical currents observed in the measurements of $I_c(H_a)$ [Fig. \ref{fig:critical-current}(inset 2)]. 
There is an abrupt transition of $V_{dc}$ to the expected sign for net vortex motion in the easy direction at  $H_a \approx \pm1$~Oe and this can be seen as vertical ridges in Figure \ref{fig:density-plot}.

By comparing line cuts of $V_{dc}(H_a)$ for a particular value of $I_{ac}$ [Fig. \ref{fig:density-plot}(inset 2)] with the measurements of $I_c(H_a)$ [Fig. \ref{fig:critical-current}(inset 3)], we observe that 
the value of $H_a$ at which $|V_{dc}|$ reaches the maximum ($H_a = 8.6$~Oe) coincides with the approximate convergence of the two senses of $I_c$. 
A rough extrapolation from the peak structure in $I_c(H_a)$ indicates that the maximum in $|V_{dc}|$ occurs approximately at the matching point of one vortex per ratchet cell.
Thus, at this point, the arrangement of vortices minimizes the asymmetry in the static friction, as characterized by $I_c$, yet the overall ratchet response, as captured by $|V_{dc}|$ is a maximum, due to the substantial asymmetry in this regime between the dynamical sliding states for the two directions. 
The two branches of the IVC measured with a large $I_{ac}$ [Fig. \ref{fig:density-plot}(inset 3)] exhibit a considerable difference in curvature and 
this dynamical asymmetry results in the significant ratchet response.


In summary, we have demonstrated a substantial ratchet effect for a system of vortices moving through weak-pinning channels with asymmetric walls.
This ratchet exhibits considerable asymmetries in both the static and dynamic friction, with different dependences on $H_a$. 
The edge barrier corresponding to the strip geometry of our structure has an important role in the vortex dynamics, including delineating a low-field Meissner regime in the channels from a state corresponding to vortices occupying ratchet cells statically for $I < I_c$. 
However, asymmetries in the edge barriers alone, as described by the model of Ref. \cite{vodolazov05}, cannot account for our ratchet effect, although this may be related to the smaller reverse ratchet response that we observe at small $H_a$ in the Meissner regime of the channels.
The microfabricated nature of our channels allows for future ratchet explorations with different channel wall shapes and configurations.


This work was supported by the National Science Foundation under Grant DMR-0547147. We acknowledge use of the Cornell NanoScale Facility.

\bibliography{vortex}

\end{document}